**Author for correspondence:**
Edward D. Lee
e-mail: edlee@santafe.edu

**THE ROYAL SOCIETY**
PUBLISHING

# Sensitivity of collective outcomes identifies pivotal components

Edward D. Lee[1,2], Daniel M. Katz[3], Michael J. Bommarito II[3] and Paul H. Ginsparg[1]

[1]Department of Physics, 142 Sciences Drive, Cornell University, Ithaca, NY 14853, USA
[2]Santa Fe Institute, 1399 Hyde Park Road, Santa Fe, NM 87501, USA
[3]Chicago-Kent College of Law, Illinois Institute of Technology, 565 West Adams, Chicago, IL 60661, USA

EDL, 0000-0003-2075-6342

A social system is susceptible to perturbation when its collective properties depend sensitively on a few pivotal components. Using the information geometry of minimal models from statistical physics, we develop an approach to identify pivotal components to which coarse-grained, or aggregate, properties are sensitive. As an example, we introduce our approach on a reduced toy model with a median voter who always votes in the majority. The sensitivity of majority–minority divisions to changing voter behaviour pinpoints the unique role of the median. More generally, the sensitivity identifies pivotal components that precisely determine collective outcomes generated by a complex network of interactions. Using perturbations to target pivotal components in the models, we analyse datasets from political voting, finance and Twitter. Across these systems, we find remarkable variety, from systems dominated by a median-like component to those whose components behave more equally. In the context of political institutions such as courts or legislatures, our methodology can help describe how changes in voters map to new collective voting outcomes. For economic indices, differing system response reflects varying fiscal conditions across time. Thus, our information-geometric approach provides a principled, quantitative framework that may help assess the robustness of collective outcomes to targeted perturbation and compare social institutions, or even biological networks, with one another and across time.

## 1. Introduction

When collective outcomes are highly sensitive to the behaviour of few individual components, these components are pivotal. Collective outcomes could be the partition of voters into blocs, the pattern of co-moving financial indices, or the coalescence of shared vocabulary in a social community. A classic example is the swing, or median, voter, prominent in political science and economics: if voters can be deterministically ranked according to preference, the median will always vote in the majority and thus is predictive of the outcome [1–3]. In real systems, this simple picture becomes much more complicated because the median might change depending on the contested issue [4], multiple issues may be at stake simultaneously [5], voters might exchange votes strategically [6,7], etc. In other words, competing interactions between voters imply that changes in individual voting behaviour may cascade into alignment or antagonistic changes in others resulting from direct physical interactions or indirect ones, i.e. mediated through a new compromise on the contents of a legislative bill. In contrast with an idealized notion of a median, we consider a 'pivotal' voter, one that could change collective outcomes even when accounting for such complexity. Here, we develop this generalized notion and use it to identify components that are especially indicative of collective changes in political voting, financial indices and social media on Twitter.

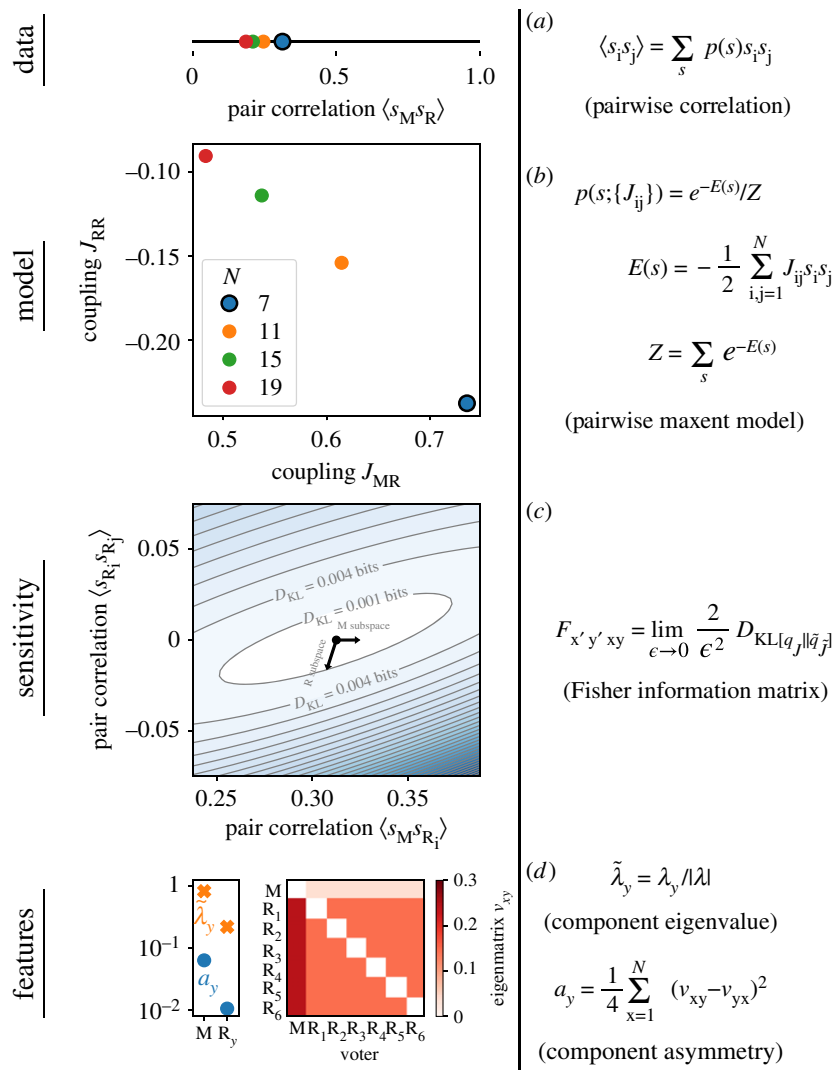

**Figure 1.** Overview of method for identifying pivotal voters for the Median Voter Model. (*a*) Taking the pairwise correlations (note $\langle s_{R_i} s_{R_{j \neq i}} \rangle = 0$), (*b*) we solve a pairwise maxent model to learn the probability distribution $p(s; \{J_{ij}\})$ parametrized by the couplings $J_{ij}$. MVMs of different sizes $N$ correspond to different coordinates in this two-dimensional space, but we focus on $N = 7$ as an example. (*c*) We calculate the FIM for $q(k)$, the probability of $k$ votes in the majority, measuring the sensitivity of $q(k)$ to changes in voter behaviour as described by equations (2.1) and (2.2). As we describe in electronic supplementary material, appendix C, the sensitivity corresponds to the curvature of the Kullback–Leibler divergence $D_{KL}$. We show a two-dimensional cut of $D_{KL}$ when M copies $R_i$ and $R_i$ copies $R_j$. The principal directions in the full space, $v_{xy}$, determine the combinations of perturbations to which $q(k)$ is most sensitive. We show projections of eigenvectors obtained from limiting perturbations to the Median or an Random voter as black arrows. (*d*) The principal eigenvector of the FIM, reshaped into an 'eigenmatrix' on the right, specifies the relative change in the rate that $x$'s votes are replaced by $y$'s (i.e. a positive value is the rate at which $x$'s voting record becomes $y$'s and a negative the rate at which its disagreements increase). The vector of principal subspace eigenvalues per voter $\tilde{\lambda}_i$, corresponding to the outlined diagonal blocks in electronic supplementary material, figure S.2C, are divided by norm $|\lambda|$ to give our pivotal measure. The asymmetry $a_y$ measures the difference in perturbations localized to a specific voter versus all its neighbours in turn. If a voter and all its neighbours are similar, the asymmetry is close to zero. Otherwise, it is bounded by a maximum value of one (see electronic supplementary material, appendix D).

Information geometry provides a natural framework for measuring how sensitive collective properties are to change in component behaviour. The Fisher information, the fundamental quantity of information geometry, establishes a unique, invariant metric over probability distributions $p(s; \{J_{ij}\})$ of states $s$ (e.g. a voting configuration, stock and sector price movement, behaviour on social networks) determined by the set of parameters $\{J_{ij}\}$ indexed by $ij$ [8]. When the parameters are infinitesimally changed to $\{\tilde{J}_{ij}\}$, the distribution becomes $p(s; \{\tilde{J}_{ij}\})$, and the distance between the two distributions is given by the Fisher information (FI) [9]. By measuring the FI for perturbations to each pair of variables $J_{ij}$ and $J_{i'j'}$ in turn, we construct the Fisher information matrix (FIM) $F_{iji'j'}$, whose eigenvectors describe how changes to the parameters lead either to sharp change in the model (large eigenvalues) or slow change (small eigenvalues)

[10,11]. In our approach, we coarse-grain states $s$ to a lower-dimensional collective outcome $f(s)$, mapping the probability distribution to one over the coarse-grained state $p(s) \rightarrow q[f(s)]$. An example is when the full vector of votes is compressed into a majority outcome. By computing the FIM over the distribution of coarse-grained states, we investigate when aggregate properties are highly sensitive to a few components, components that we determine to be pivotal to the system's collective properties.

We outline our approach in figure 1, where we fit a minimal statistical model to a dataset, measure the FIM and extract properties of the local information geometry. We first discuss this approach on a toy median voter model to build intuition, and we extend detailed analysis to voting on an example from the US Supreme Court (SCOTUS) and State Street Global Advisors SPDR exchange-traded funds [12,13].

Then, we perform a survey across multiple systems in society including examples of judicial voting across US state high courts [14], California (CA) state legislatures [15], US federal legislatures [16], and communities on Twitter [17]. Across these examples, we find large diversity ranging from examples of median-like systems, with pivotal components, to other examples in which no special component emerges.

## 2. Results

### 2.1. Median Voter Model

The role of the median derives from the fact that in a majority-rule voting system, the voting outcome is a coarse-graining instead of depending on the detailed nature of every individual's vote. The margin by which the majority wins, as is captured in the probability $q(k)$ that $k$ voters of the system are in the majority, can reflect the appeal of the voting outcome or even its legitimacy, perceptions of which feed back into the decision process [18]. Thus, $q(k)$ serves as an aggregate measure of underlying decision dynamics that we will use to identify pivotal blocs.

To outline our approach, we study the sensitivity of $q(k)$ in the context of a reduced toy model that captures the essence of a median voter. The ideal median voter exists in a majority-rule system where voters' preferences are unidimensional. By virtue of the unique ranking of preference, the median is always in the majority [1]. We propose a statistical generalization, the Median Voter Model (MVM), with an odd number of $N$ voters. The MVM consists of $N-1$ Random voters and one Median voter who always joins the majority. The binary vote of voter i, $s_i$, is equally likely to be $-1$ and $1$ such that only majority-minority divisions are relevant. As a result, the average votes are all the same, but the set of pairwise correlations as shown in figure 1a for $N=7$ voters are nonzero between M and R, $\langle s_M s_R \rangle = 0.3125$, and zero between R's, $\langle s_{R_i} s_{R_{j \neq i}} \rangle = 0$. Thus, this model consists of a special voter, the Median (M), who after a voting sample has been taken, is perfectly correlated with the majority, whereas Random (R) voters all behave in a statistically uniform and random way.

To capture the network of interactions between individuals from which majority-minority coalitions emerge, we take a pairwise maximum entropy (maxent) approach [19]. The maxent principle describes a way of building minimal models based on data. We maximize the information entropy $S = -\sum_s p(s) \ln p(s)$ while fixing the model to match the pairwise correlations from the data, $\langle s_i s_j \rangle_{\text{data}} = \langle s_i s_j \rangle$ as defined in figure 1a. The result is a minimal model parametrized by *statistical* interactions between voters, or 'couplings' $J_{ij}$ in figure 1b [20]. For each pair of voters with pairwise correlations in figure 1a, there is a corresponding coupling such that the set of couplings is specified exactly by the pairwise correlation matrix. For the MVM, the $\binom{N}{2}$ couplings only take two possible values, one for each of the two unique correlations. The couplings for the MVM indicate that all R's tend to vote with M (agreement between M and R leads to an increase in the log-probability $\ln r(s_M = s_R) \propto J_{MR}$ as in figure 1b) with a slight tendency for R's to disagree with each other more than would be expected given their shared correlation with M (disagreement between $R_i$ and $R_j$ decreases the log-probability of the vote by $\ln r(s_{R_i} = s_{R_j}) \propto J_{RR}$). In principle, any probabilistic graph model is a viable alternative for the approach we outline, but the pairwise maxent model has been shown to capture voting statistics better than other models of voting with surprisingly few parameters [21,22], fits the data well (electronic supplementary material, appendix B), and presents a particularly tractable formulation for calculating information quantities.

To probe how the collective properties captured by the distribution $q(k)$ depend on the voters, we ask how the distribution would change if the voters were slightly different. In this example of majority-minority voting, any change in voting behaviour is reflected in the pairwise correlations and preserves the symmetry between the two possible outcomes $-1$ and $1$. A natural endpoint for the set of possible $q(k)$ as we increase the pairwise correlations is when all voters are perfectly correlated, so we consider perturbations that take us towards this endpoint: with probability $\epsilon$, voter y's votes are replaced by x's,

$$\bar{r}(s_y = s_x) = (1 - \epsilon)r(s_y = s_x) + \epsilon. \tag{2.1}$$

Equation (2.1) is a weighted average that interpolates from the observed probability of agreement between x and y when $\epsilon = 0$ to perfect agreement when $\epsilon = 1$. We then account for the changes to y's correlations with the remaining voters:

$$\bar{r}(s_y = s_{x' \neq x}) = (1 - \epsilon)r(s_y = s_{x'}) + \epsilon r(s_x = s_{x'}). \tag{2.2}$$

Equation (2.2) interpolates from the observed probability of agreement between y and x' when $\epsilon = 0$ to that between x and x' when $\epsilon = 1$. If replacing M with any R voter such that $y = M$ and $x = R$, the operation defined in equation (2.1) increases the pairwise correlation $\langle s_M s_R \rangle$ while simultaneously changing M's correlations with the others to be more like those with R, pushing them to zero. When the statistical model exactly matches the entire distribution of votes $p(s) = p_{\text{data}}(s)$, the perturbation described in equations (2.1) and (2.2) is equivalent to shifting the probability from any voting configuration where i and j disagree to the voting configuration where i and j agree, holding all other probabilities constant. With the pairwise maxent model, however, the perturbation is only reflected in the pairwise correlations, moving us from one model to another within the class of pairwise maxent models. In this case, the perturbations can be mapped to changes in the couplings $J_{ij}$ in the limit of $\epsilon \to 0$ that we use to determine the entries of the FIM shown in electronic supplementary material, figure S.2C (electronic supplementary material, appendix C).

The variation in the entries of the FIM indicates the unique role of the median. The FIM describes the curvature of the Kullback-Leibler divergence $D_{KL}$ as the probabilities of pairwise agreement are modified. Under small perturbations, the contours of $D_{KL}$ form an ellipse, whose major and minor axes represent components of the FIM's eigenvectors, as in figure 1c. We show the principal eigenvector in figure 1d. Its entries represent the relative amount by which pairs should be simultaneously varied for maximal local change to $q(k)$—as if one could change all the pairwise voting 'knobs' at once. To be clear about the pairwise grouping of index, we reshape the principal eigenvector into an 'eigenmatrix' in figure 1d. Each column corresponds to a directed change where voter y is made more similar to the corresponding row voter x. Since R's are all the same, the first column connecting M to each R is uniformly valued. In the first row, the entries all correspond to making the neighbours of M more like M, so these are also all uniformly valued given that the R's are interchangeable. Thus, each column of the eigenmatrix describes perturbations localized to the column voter and each row corresponds to changes across all the neighbours of a particular voter such that the symmetry between R's and the unusual role of M manifests in the comparison of local neighbourhood with the local neighbourhood of neighbours.

This local versus neighbourhood asymmetry presents one way of pinpointing an unusual voter by using the difference between the eigenmatrix $v_{xy}$ and its transpose $v_{yx}$. We define this per voter asymmetry $a_y$ in figure 1$d$. Given a normalized eigenmatrix, the total asymmetry over all voters $A \equiv \sum_y a_y$ is 0 when the eigenmatrix is perfectly symmetric and is 1 when perfectly antisymmetric $v_{xy} = -v_{yx}$. The point $A = 1/2$ marks the maximum asymmetry possible when all the nonzero elements are of the same sign—such as when for each $v_{xy} > 0$, $v_{yx} = 0$ (electronic supplementary material, appendix D). For the MVM with $N = 7$, we find that M's asymmetry $a_M = 0.06$, whereas $a_R = 0.01$, clearly distinguishing M from R. The total $A = 0.13$, a point of reference for systems that are more complex than the MVM. For larger $N$, the MVM asymmetry $a_M$ grows as the role of M more visibly skews the distribution. Thus, both the asymmetry in the roles of voters and the growing importance of a median with system size are reflected in the symmetry of the eigenmatrices.

To measure the sensitivity of $q(k)$ to each voter, we inspect the subspace eigenvalues $\lambda_i$, specifying the sensitivity of $q(k)$ to change in a single voter's behaviour. These values are calculated from the subspace of the FIM describing localized perturbations—the diagonal blocks of the FIM as outlined in electronic supplementary material, figure S2C, and whose eigenvectors are projected into figure 1$c$. The upper leftmost block of the FIM corresponds to M and the remaining blocks correspond to each R in turn. For each subspace, we retrieve the principal eigenvalue. To compare the eigenvalues across voters, we calculate the normalized eigenvalue as defined in figure 1$d$, defining our measure of how 'pivotal' a component is relative to others. For the $N = 7$ MVM, the principal eigenvalues are $\tilde{\lambda}_M = 0.70$ and $\tilde{\lambda}_R = 0.05$. This large difference indicates that $q(k)$ is over 10 times more sensitive to variation in M than R, again reaffirming the special role of the median. It is important to note that voters with strong asymmetry are not necessarily the most pivotal—clearly because eigenvalues and eigenvectors present different information. Still, asymmetry in the eigenmatrix indicates heterogeneity among the voters; thus, large asymmetry is necessary, if insufficient, for the pivotal measure to vary across a wide range. Overall, the information geometry of this minimal class of models provides a way of quantifying the role of individual components on collective outcomes, identifying key components with pivotal roles that can emerge given strong heterogeneity in the population.

## 2.2. US Supreme Court and S&P 500

We perform the same analysis on an example from SCOTUS of $N = 9$ voters, $K = 909$ votes, and between the years 1994 and 2005 (see §4.3 for details about datasets). We show the principal eigenmatrix in figure 2 that consists of perturbations primarily increasing similarity across ideological wings given by the positive values connecting liberals and conservatives.[1] The principal mode has a total asymmetry of $A = 0.10$ compared to $A = 0.25$ for the $N = 9$ MVM, indicating the absence of a median-like, pivotal voter. This absence is surprising because discussion of medians A. Kennedy and S. O'Connor is prominent in the context of this court. When we consider voter-subspace eigenvalues shown in figure 2, we find the justices in ranked order: C. Thomas, S. Breyer and Chief Justice W. Rehnquist. A change in C.T., given his strongly conservative voting record, would naturally constitute consequential change, but the roles of W.R. and S.B. are more subtle [21,24,25]. Despite A.K. and S.O.'s prominent role in the narrative of Supreme Court voting, we find that other justices come to the foreground when we consider the sensitivity of the Court to behavioural change.

The principal mode can be projected into the more intuitive space of dissenting coalitions in terms of the rate of change of the probabilities for dissenting blocs (§4.2). Though the eigenmatrix in figure 2 shows increasing similarity between ideological wings, suggesting suppression of partisan 5–4 divides, the frequency of any 5–4 divide actually increases strongly along with a decrease in lone and pair dissents as in the bottom of figure 2. Seven of the nine most common pair dissents found in the data decrease in likelihood. Thus, this shift reflects an increasing tendency for justices to join larger blocs, reflected in the suppression of every justice's lone dissents in a way that breaks the typical partisan divide. To visualize changes in the existing 5–4 conservative–liberal dynamic, we inspect defections from the liberal bloc, or 6–3 votes where a single liberal vote is missing, and likewise defections from the five-member conservative bloc. On the whole, defections from the liberal bloc are less surprising than those for the conservative bloc, consistent with the balance of power favouring conservatives. For the liberal bloc, the most prominent change entails R.G. defecting, leaving D.S., J.S., S.B., which reflects the central role of R.G. in the liberal coalition. On the other side, increasing the probability of S.O. or A.K. defecting is important though not as much as the defection of W.R., which reflects his often-understated, unusual statistical role in the Court [21]. Consistent with pundits' understanding is the large surprise associated with C.T.'s defection from the conservative majority, a change that would represent a fundamental shift in the established partisan dynamics. Overall, this individual variation in the context of the partisan 5–4 dynamic reveals a portrait of much deeper subtlety than that suggested by unidimensional partisan intuition [4,21,26]. Thus, the information geometry of statistical models of social systems can provide detailed insight into specific components or blocs in direct connection to their role in collective modes of the system.

In figure 3, we analyse the founding set of State Street Global Advisors SPDR exchange-traded funds ($N = 9$; $K = 4779$; 2000–2018), which replicate the indices and provide daily price data (binarized to positive $s_i = 1$ or negative daily changes including no change $s_i = -1$ in analogy to votes). In contrast with SCOTUS, the collective behaviour of each index reflects the aggregation of many individual investors: no stock index is monolithic in the sense of an individual voter. Given this aggregate nature, it is natural to consider the eigenvectors as the most surprising set of unanticipated global changes—although entire sectors might be 'perturbed' by government policy like sector-specific regulation or tariffs. From this point of view, fluctuations in the pivotal blocs might reveal notable shifts in economic conditions or collective perceptions thereof (electronic supplementary material, appendix E). Taking a look at the model, we find that the principal mode displays large asymmetry across every index, reflecting the diversity of roles played by the various sectors of the economy as captured in price movements. Relatively large subspace eigenvalues highlight XLE (energy) and XLU (utilities), in agreement with their role as drivers of the economy on whose outputs many of the other sectors depend [27,28]. Perhaps unsurprisingly, we also find a 'bellwether' XLP (consumer staples) and XLV (healthcare) as notably pivotal whereas XLF (financial) and XLI (industrials) seem to be relatively not. Going beyond the principal mode, we inspect the secondary mode and find that it is remarkably symmetric, with an asymmetry score of $A = 0.07$, in contrast with the second mode of the SCOTUS example where $A = 0.44$. This secondary symmetry is reminiscent of the MVM where a prominent asymmetric mode hides a nearly symmetric mode arising from the uniformity of Random voters. Such a symmetry is not found for the SCOTUS example, where at lower modes, asymmetry actually increases, signalling notable individual roles in determining collective outcomes. Taken together, these examples are a comparison of opposites, where the apparent asymmetry in components obscures shared structure for SPDR, whereas for SCOTUS the overarching tendency to consensus overshadows individual roles on the Court.

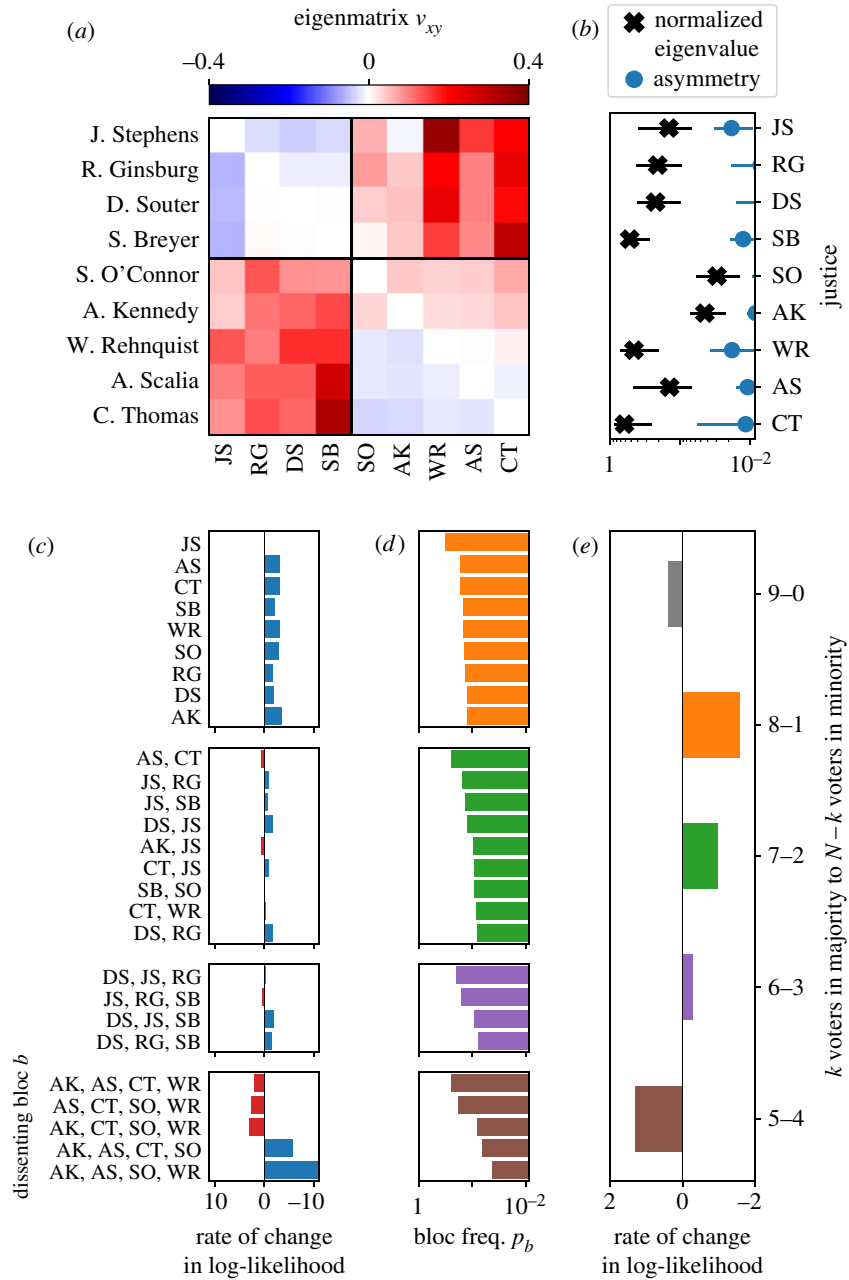

**Figure 2.** SCOTUS example. (*a*) Principal eigenmatrix of the FIM. Justices are ordered from most liberal to most conservative according to a standard measure of ideology [23]. We indicate the typical divisions between the liberal and conservative blocs with black lines. (*b*) Normalized voter-subspace eigenvalues and asymmetry per justice as defined in figure 1*d*. (*c*) Rate of change in log-probability of dissent for dissenting blocs ln $p_b$. (*d*) Each bloc *b*'s probability of dissenting together according to the pairwise maxent model, $p_b$. (*e*) Rate of change in log-probability of *k* dissenters ln $q(k)$. Error bars represent 95% confidence intervals from repeating the full procedure outlined in figure 1 for $10^2$ bootstrapped samples of the data.

## 2.3. Pivotal components in society

We explore other examples of social systems, including votes from US state high courts [14], the California State Assembly and Senate [15], the US federal legislature [16], and communities on Twitter [17]. As with the previous two examples, we map behaviour in these systems to binary form. To reduce the larger legislative bodies to a comparable number of blocs, we first separate voters into 9 nearly-equally-sized blocs by ranked similarity according to a standard political science measure of ideology, the first W-Nominate dimension [29]. The bloc vote is given by the majority vote of the members and is randomly chosen if equally divided (see electronic supplementary material, figure S.14). For Twitter communities, we identify individuals as high-dimensional binary vectors where an element is positive if they used a corresponding keyword, or else negative, such that the pairwise correlations reflect overlap in their use of keywords. As outlined above, our analysis of the information geometry involves the same procedure but for a wider variety of social systems.

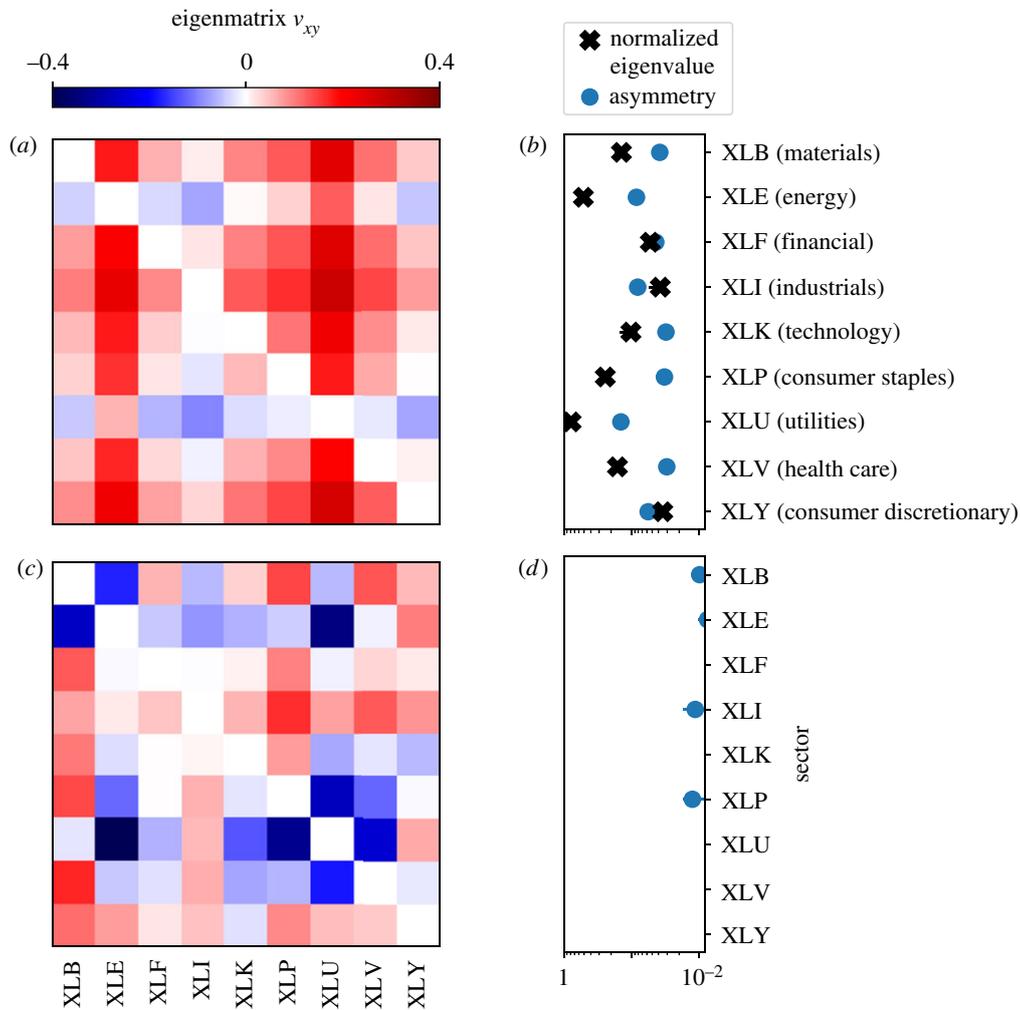

**Figure 3.** S&P SPDR example. (*a*) Principal and (*c*) secondary eigenmatrices of the FIM with eigenvalues $\Lambda_1 = 0.41$ and $\Lambda_2 = 0.13$, respectively. (*b*,*d*) Relative sector subspace eigenvalues and asymmetry by sector (figure 1*d*). Error bars represent 95% bootstrapped confidence intervals.

Considering the principal eigenmatrix of the Alaskan (AK) Supreme Court ($N = 5$; $K = 1021$; 1998–2007), we find a remarkable degree of symmetry between justices and a small value for the total asymmetry $A = 0.01$. Such symmetry implies that the justices on this court all dissent in a statistically uniform way as described by the set of their pairwise correlations. Though this could be trivially true if all pairwise couplings were the same, this is not the case, a fact that is mirrored in the spread of positive and negative values in the eigenmatrix in figure 4. Checking the local interaction networks described by the set of couplings to every neighbour j for justice i (electronic supplementary material, figure S.3), we find that the sets are all similar for every justice i. This symmetry is mirrored in the similarity of the individual subspace eigenvalues shown in figure 4. Consistent with this symmetry extracted from the voting record, four out of the five justices served as Chief Justice during this period,[2] a regular rotation of roles imposed by the state constitution stipulating that the Chief Justice only serve for three consecutive years at a time. In contrast, we show that the New Jersey (NJ) Supreme Court ($N = 7$; $K = 185$; 2007–2010) has strong asymmetry of $A = 0.5$ (electronic supplementary material, figure S.8). Appointments to the NJ Court follow a tradition of maintaining partisan balance, apparently codifying a median role into the institution, and we find two nearly equal pivotal voters. Despite the seeming alignment between each of these two examples and the institutional norms, AK Supreme Courts are not always less symmetric than their NJ counterparts. The asymmetry is highly variable for previous years, suggesting that codified institutional rules only partially determine the role of pivotal voters (electronic supplementary material, figure S.15).

We also show the eigenmatrices of the 1999 session of the CA State Assembly ($N = 77$; $K = 5424$; 1999–2000) and a K-pop Twitter community ($N = 10$; $K = 7940$; 2009–2017). The CA Assembly is an example of strong asymmetry ($A = 0.46$). For the sessions starting between the years 1993 and 2017, we find that the Assembly displays stronger signatures of asymmetry (average total asymmetry $\langle A \rangle = 0.4 \pm 0.1$) compared to the Senate ($\langle A \rangle = 0.3 \pm 0.1$), showing how the rules of the institution might be reflected in the distribution of pivotal blocs. We then compare the distribution of the single largest pivotal measure $\tilde{\lambda}_{max}$ with that of the similarly coarse-grained US House of Representatives and Senate. Though the CA distributions for $\tilde{\lambda}_{max}$ are statistically indistinguishable from each other (Kolmogorov–Smirnov test statistic $k = 0.31$ and significance level $p = 0.5$) and the federal bodies' distributions are similar ($k = 0.31$, $p = 0.05$), the state versus federal levels show larger differences ($k > 0.46$, $p < 0.03$). This separation between the behaviour at state and federal legislatures reflects institutional differences that are captured in the sensitivity of majority–minority coalitions (see electronic supplementary material, figure S.12).

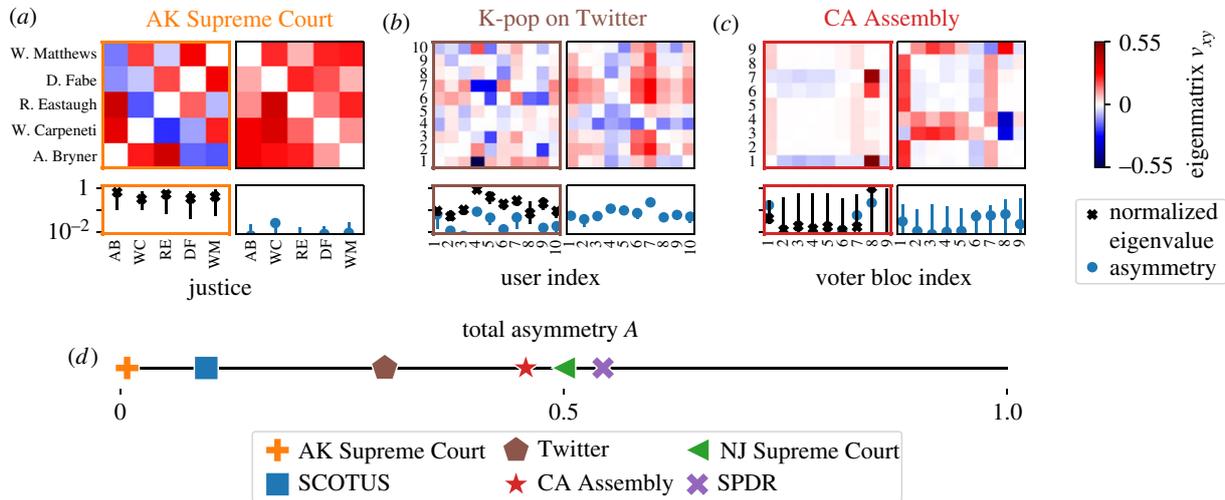

**Figure 4.** Example systems. (*a*–*c*) We show the first two eigenmatrices for AK Supreme Court, K-pop on Twitter, and CA Assembly. Below each eigenmatrix, we show the pivotal measure and the asymmetry per component as defined in figure 1*d*. For the CA Assembly, voters are grouped into nine blocs after rank ordering by the first W-Nominate dimension (see electronic supplementary material, figure S.14). Large error bars in the pivotal measure indicate that the unique pivotal role of Bloc 8 depends on a few crucial votes in this session. (*d*) Asymmetry of the principal eigenmatrix of the FIM. Error bars represent 95% bootstrapped confidence intervals.

As for the Twitter K-pop community, we find much heterogeneity among users with total asymmetry $A = 0.30$, exceeding that of the $N = 9$ MVM. In contrast with the MVM, this community contains multiple pivotal members but wide variation in the strength of their subspace eigenvalues. Twitter communities may be on average sensitive to the behaviour of a few individuals [17], but this individual-level variation suggests that collective behaviour may be much more sensitive to a select Twitter users even within smaller communities [30]. Going beyond the detailed few examples in figure 4, we find large diversity within political institutions that highlights the important role of heterogeneity in social institutions, heterogeneity that is captured in the information geometry of minimal, maxent models.

## 3. Discussion

An important question in the study of social institutions is whether or not collective decisions are robust to perturbation targeting individual components. Robustness is reciprocal to sensitivity: when a system is highly sensitive to small changes to components, its collective properties are not robust. In neural networks with avalanches of firing activity [31–33], in bird flocks with propagating velocity fluctuations [34], or in macaque societies with conflict cascades [35], such sensitivity might have an adaptive functional role. In the context of human society, questions of robustness are relevant to the stability of voting coalitions or the susceptibility of a population to disease or disinformation. For example, we might be interested in comparing the impact of different judicial nominees on the dynamics of voting on a judicial bench or the impact of modified user behaviour on the spread of disinformation in social networks. By relying on the formal framework of information geometry to investigate statistical signatures of sensitivity, we present a data-driven and general approach to characterizing robustness. As a result, our approach is not model-specific, only relying on the calculation of how sensitive a model is to changes in observable individual behaviour as summarized in figure 1.

In voting systems, median voters are conventionally considered to be power brokers who have outsize influence [1,7,23]. Building on this idea, we propose a reduced toy model to extract features of a median voter. We show how to identify and interpret signatures of strong sensitivity on individual components in multiple social contexts, generalizing the intuition behind the median to pivotal components on which aggregate properties, measured by majority–minority divisions, depend strongly. Intriguingly, we find hints that institutional differences may contribute to structuring individual roles in collective outcomes both in courts and legislatures. Though it is unsurprising that the particular rules of a voting body may structure bloc dynamics, pivotal components provide a principled way of comparing social systems with differing composition, from different eras, and across different institutions in a unified, quantitative framework.

In social choice theory, the question of how institutional rules structure outcomes has been long-studied, especially in the context of political science [1,36,37]. We examine the collective structure of the system expressed in its statistics, an approach closer to spatial voting analysis [4,21,23,25,26,38,39] and complementary to one starting from first principles and then determining resulting constraints on outcomes (such as Arrow's impossibility theorem [3,40]). From such examination, we propose how one could extract the underlying levers on which collective outcomes pivot. In this sense, our framework provides a way for investigating collective decisions in systems where the 'voting rules' are unknown—even in biological systems, where the analogy to voting rules is looser but information is likewise encoded in collective states [31,41].

There are several ways by which our work might be extended to questions in the realm of social choice theory. One possibility is to analyse how different strategies (such as in tactical voting [6]) or different rule sets change the statistics of outcomes. This question could be turned around and posed as one about inferring underlying rules when they are unknown and with limited data on outcomes, a natural framework for considering how players infer the rules by which others are playing [42,43]. Though we limit ourselves to analogues of majority voting outcomes in a binary voting system, our framework is easily generalizable to other types of voting. For example, a straightforward extension would be to consider multiple outcomes. In such scenarios, a common technique for aggregating votes is the Borda count, or ranked choice, where voters rank their top choices, and the most favoured outcome is selected [44–46]. One generalization of our MVM to this situation would be to have a voter who always agrees on the top outcome and with the favoured outcomes thereafter. A final variation could be to measure how the degree of pivot varies when different issues are up for vote in political bodies. In this final scenario, it may be interesting to consider how a pivotal component could facilitate or diminish satisfaction of collective preference under measures like Pareto efficiency [47]. With variations like these on the space of possible outcomes and aggregation of preference, we propose a statistical framework for considering how collective outcomes are maximally or minimally susceptible to voting system properties. Overall, our work provides an opportunity to leverage a statistical, information-geometric framework to answer intriguing, and perhaps new, questions about how institutions structure outcomes both in data and theory.

More intuitively, we might think of pivotal components as 'knobs' that could drive a system out of its current configuration described by the ensemble $p(s)$. If the subspace eigenvectors are knobs, the pivotal measure is inversely proportional to the spacing of the dials such that for large eigenvalues the smallest turn results in the strongest effect. Since each pivotal component only considers the effects of perturbations localized to a single component, these knobs are independent. If $n$ components were accessible simultaneously, however, we would consider the joint space of multiple pivotal components, and the principal subspace eigenvalue must increase beyond (or stay at) the maximum eigenvalue over the set of component subspaces: this reflects the fact that enhancing the breadth of control only increases the range of possible outcomes [48,49]. By considering which knobs are accessible experimentally, our analysis could be extended to measuring signs of statistical control in real systems. For judicial voting, the realizable knobs that change judicial voting behaviour may be the submission of *amicus curiae* briefs, choice of litigating cases, or lobbying.[3] Those trying to craft a legislative coalition might 'perturb' aspects of proposed policy to affect its acceptability to potential supporters [50]. In controlled biological systems, localized perturbations to single components could include manipulation of single neurons or the upregulation of specific genes.[4] Our work presents the possibility of informing the direction of such external perturbations in the broader context of control.

The understanding of the interplay between components and multi-component structures across social and biological examples remains nascent at mesoscopic and macroscopic scales. With this principled, quantitative approach for measuring pivotal components, we might, by comparing systems, better understand how institutional and environmental factors shape the emergence of social structure.

# 4. Methods

## 4.1. Calculation of the Fisher information matrix

Here, we calculate the FIM for the transformation described in equations (2.1) and (2.2) and go through some examples to show how to calculate the FIM. We go into some detail with the derivation to make clear how to perform such a calculation for those less familiar with maxent models and information geometry.

In equations (2.1) and (2.2), we consider how the correlations between component y and all other components x′ change when component y appears to vote more like component x. To effect this perturbation, we use a parameter $\epsilon \to 0$ that leads to a linear change in the couplings $J_{x'y}$ as described by the rate of change $dJ_{x'y}$, where we are taking a total derivative with respect to the change in the pairwise probabilities described by the vector $r(s_x = s_y)$. To obtain this derivative, we perturb to first order in $\epsilon$ the expression for the pairwise correlations (equation (S.2)) to obtain the self-consistent equation

$$\langle s_x s_y \rangle_{\text{pert}} - \langle s_x s_y \rangle = \frac{1}{2} \sum_{x',y'}^{N} \Delta_{xy} J_{x'y'} \left( \langle s_x s_y s_{x'} s_{y'} \rangle - \langle s_x s_y \rangle_{\text{pert}} \langle s_{x'} s_{y'} \rangle \right). \tag{4.1}$$

By self-consistent, we are referring to the fact that the new pairwise correlations after perturbation $\langle s_i s_j \rangle_{\text{pert}}$ depend on the change in the couplings $\Delta_{xy} J_{x'y'}$, so the perturbations to the couplings determine quantities on both sides of equation (4.1).[5] The coupling perturbations, $\Delta_{xy} J_{x'y'}$, are related to the linear response of the couplings to change in the collective statistics induced by perturbing the pair of components x and y,

$$dJ_{x'y'} = \lim_{\epsilon \to 0} \Delta_{xy} J_{x'y'} / \epsilon. \tag{4.2}$$

The resulting matrix of new couplings is

$$\tilde{J}_{x'y'} = J_{x'y'} + \Delta_{xy} J_{x'y'} \tag{4.3}$$

and

$$\tilde{p}(s; \{\tilde{J}_{x'y'}\}) = p(s; \{J_{x'y'} + \Delta_{xy} J_{x'y'}\}). \tag{4.4}$$

The set of perturbations $\Delta_{xy} \equiv \{\Delta_{xy} J_{x'y'}\}$ are also defined in electronic supplementary material, figure S.2C.

Equations (4.1)–(4.4) describe the numerical algorithm for calculating the changes in the statistics of the system under the perturbation described in equations (2.1) and (2.2). Note that the algorithm implicitly depends on $\epsilon$, which must be taken to zero. The remaining calculations are to coarsen the full distribution $p(s)$ to $q(k)$, the distribution of $k$ votes in the majority and to calculate the FIM on $q$. For pedagogical clarity, we will first show how to calculate the FIM without coarse-graining.

There is a simple, intuitive form for the FIM for maxent models. Under an infinitesimal change in the parameters such that the energy of each voting configuration $E(s) \to E(s) + \Delta E(s)$, we can expand

$$\tilde{p}(s) = p(s)[1 + \langle \Delta E \rangle - \Delta E(s)] + \mathcal{O}(\Delta E^2). \tag{4.5}$$

Now calculating the Kullback–Leibler divergence to second order,

$$D_{KL}[p\|\tilde{p}] = \frac{1}{2}\langle(\Delta E - \langle\Delta E\rangle)^2\rangle + \mathcal{O}(\Delta E^3). \qquad (4.6)$$

The constant term is zero ($D_{KL}[p\|p]=0$) and the linear term is zero because the changes to the probability function under perturbation must sum to zero to preserve normalization. The FI is the second order term, or the curvature, so we must send the norm change in the energy $|\Delta E| \to 0$,

$$F = \lim_{\epsilon\to 0}\epsilon^{-2}\left(\langle\Delta E^2\rangle - \langle\Delta E\rangle^2\right), \qquad (4.7)$$

where $\epsilon$ is defined in equation (4.2). For the symmetrized pairwise maxent model class, $\Delta E$ is the sum over all couplings. Thus, the FI of the distribution over the full state space $p(s)$ has a simple form in terms of the change in energy for maxent models.

Alternatively, the result from equation (4.7) can be expressed as a matrix of correlation functions. In other words, the correlation functions are the linear response functions for perturbations to the natural parameters, here the couplings $J_{xy}$. As the simplest example, consider an Ising model under perturbation to a particular coupling $J_{xy}$. Using our form equation (4.7) for the FI and $\Delta J_{xy} \equiv \tilde{J}_{xy} - J_{xy}$,

$$F_{xy xy} = \lim_{\Delta J_{xy}\to 0}\Delta J_{xy}^{-2}\langle(\Delta J_{xy}s_x s_y - \Delta J_{xy}\langle s_x s_y\rangle)^2\rangle. \qquad (4.8)$$

The perturbations to the couplings $\Delta J_{xy}$ do not depend on the state $s$, so they can be pulled out of the averages to obtain

$$= \langle s_x^2 s_y^2\rangle - \langle s_x s_y\rangle^2 \qquad (4.9)$$

$$= 1 - \langle s_x s_y\rangle^2. \qquad (4.10)$$

The diagonal entries of the FIM are the variance of the pairwise correlation, which is a well-known result. It is straightforward to see that the off-diagonal elements of the FIM are the covariance $\langle s_x s_y s_{x'} s_{y'}\rangle - \langle s_x s_y\rangle\langle s_{x'} s_{y'}\rangle$. Thus, the FI for maxent models reduces to the covariance of the set of observables chosen as constraints, when we are dealing with natural parameters.

As a more general formulation, consider the set of Lagrangian multipliers $\theta_n$ and their corresponding bare observables $f_{\theta_n}(s)$ ('bare' referring to the fact that we have yet to dress them with brackets by averaging over the ensemble). For the pairwise maxent model, the Lagrangian multipliers are the couplings and the bare observables are the pairwise products $s_x s_y$. Working through the same calculation as before but with this general formulation of a maxent model, we find for the FI,

$$F = \lim_{\epsilon\to 0}\epsilon^{-2}\left\{\sum_n \Delta\theta_n^2[\langle f_{\theta_n}^2\rangle - \langle f_{\theta_n}\rangle^2]\right.$$
$$\left. + \sum_{n,m}\Delta\theta_n\Delta\theta_m[\langle f_{\theta_n}f_{\theta_m}\rangle - \langle f_{\theta_n}\rangle\langle f_{\theta_m}\rangle]\right\}, \qquad (4.11)$$

where the $\Delta\theta_n$ depend implicitly on $\epsilon$. As noted earlier, the perturbation in the pairwise agreement probabilities leads to a nontrivial combination of changes to the entire vector of couplings. As a result, the FI in equation (4.11) contains cross terms between all pairwise correlations and the change in the Lagrangian multipliers $\Delta\theta_n$ each come with a factor of the Jacobian relating changes in the pairwise marginals to the couplings as described by equation (4.1).

For the analysis in the main text, however, there is an additional step. We do not consider the full state space, but coarse-grain each $p(s)$ to the distribution of $k$ votes in the majority $q(k)$. As a result, we are not calculating the variance in the energies for the pairwise maxent model as described in equation (4.7), but the variance in the logarithm of the sum of all terms in the partition function with $k$ voters in the majority. We label the set of all states with $k$ voters in the majority $S_k$ to write

$$q(k) = \sum_{s\in S_k}p(s) = \frac{1}{Z}\sum_{s\in S_k}e^{-E(s)} = \frac{1}{Z}e^{-E_{maj}(k)} \qquad (4.12)$$

and

$$E_{maj}(k) \equiv -\ln\left(\sum_{s\in S_k}e^{-E(s)}\right). \qquad (4.13)$$

Equation (4.13) defines an effective '$k$ majority' energy such that under perturbation to the pair of components x and y as indicated by $\Delta_{xy}$

$$F_{xy} = \lim_{\epsilon\to 0}\epsilon^{-2}\left(\langle\Delta_{xy}E_{maj}^2\rangle - \langle\Delta_{xy}E_{maj}\rangle^2\right). \qquad (4.14)$$

Equation (4.14) is the form that the limit in figure 1c takes.

To summarize the algorithm, we first find the total derivative of each coupling $dJ_{x'y'}$ with respect to the change in the pairwise marginals as explained in equation (4.2). Then, we calculate the change in the distribution of $k$ votes in the majority for both the model without the perturbation $q(k)$ and with $\tilde{q}(k)$ for a range of small values $\epsilon$ as in equation (4.6). By comparing these two distributions for increasingly smaller $\epsilon$, we estimate numerically the FIM, relying on the definition of an 'effective' energy $E_{maj}$ as in equation (4.13) to deal with issues in numerical precision that may arise when comparing ratios of floating point numbers. These steps generate the FIM as shown in figure 1c with which we calculate the eigenvalue spectrum to measure our pivotal voters.

## 4.2. Dissenting coalitions

In figure 2, we project the eigenvectors onto the probabilities of dissenting coalitions to obtain a detailed picture of how the parameter directions obtained from the FIM affect dissenting coalitions. Such a projection involves taking the sum over all the probabilities of the states with the particular dissenting bloc and calculating the effective energy. Expanding the log-likelihood to first order, we calculate the rate at which this probability changes to be

$$d\ln q(k) = \lim_{\epsilon\to 0}\frac{1}{\epsilon}\left(\Delta E_{maj}(k) - \langle\Delta E_{maj}(k)\rangle\right). \qquad (4.15)$$

The limit $\epsilon \to 0$ refers to an infinitesimal perturbation of $q(k)$ along the first eigenvector of the FIM. Then, the rate of change in the log-likelihood simplifies to comparing the change in the effective majority energy $E_{maj}(k)$ with the average change across $p(s)$.

## 4.3. Additional notes on datasets

### 4.3.1. US state supreme courts

We obtained the latest dataset from the State Supreme Court Data Project (SSCDP) and used their binary coding of justice votes [14].

We show the total asymmetry for all the natural courts on the Alaska and New Jersey Supreme Courts we considered in electronic supplementary material, figure S.15. We only considered natural courts with at least 100 where the full complement of justices were voting. As we mention in the main text, there is variation in the measured value of asymmetry that makes unclear relationship between total asymmetry and codified institutional rules of voting.

### 4.3.2. US Supreme Court

We use data from the Supreme Court Database Version 2016 Release 1, taking their binary coding of majority–minority votes [12]. This same dataset and version has been analysed previously (see [21,22]).

### 4.3.3. SPDR
The SPDR Select Sector indices track the Standard & Poor's (S&P) and Morgan Stanley Capital International (MSCI) Global Industry Classification Standard (GICS) sectors. As described on Wikipedia, GICS 'is an industry taxonomy developed in 1999 by MSCI and S&P for use by the global financial community. The GICS structure consists of 11 sectors, 24 industry groups, 69 industries and 158 sub-industries into which S&P has categorized all major public companies. The system is similar to the Industry Classification Benchmark, a classification structure maintained by FTSE [Financial Times Stock Exchange] Group.'

We focus on these assets and their adjusted price action because (i) they are the most heavily traded and representative sector assets in the world, so their prices and volumes reflect actual interest in exposure to the sectors, (ii) they have been traded daily without exception for over 20 years, and (iii) unlike the Dow indices, the S&P indices are not subject to effects of price-weighting such as reverse-split over-weighting. The historical price data are available online on Yahoo! Finance.

### 4.3.4. Twitter
We analyse one of the communities from the data considered in [17]. In this work, the authors divide the Twitter community into smaller subcommunities using the CNM algorithm [51]. We take one example from their K-pop community with 10 individuals.

### 4.3.5. CA Assembly and Senate
Session records were obtained from Prof. Jeff Lewis' scrape of the CA legislature's public data API [15]. For all sessions from 1993 to 2017, we solved the W-Nominate model using the code provided in [29]. We then removed any voter who did not participate in more than 20% of the votes, rank-ordered the voters by the first W-Nominate dimension, and divided them as equally as possible into 9 groups as shown for the 1999–2000 session in electronic supplementary material, figure S.14.

For the results of bootstrap sampling to calculate error bars, we found that 3% of the Assembly samples showed significant error from the fit correlations because of numerical precision issues. This is generally an issue for systems that are poised near the boundaries of the model manifold where the couplings become large. For the error bars on the normalized subspace eigenvalues, however, the contribution from these three missing samples is negligible.

In figure 4, the most pivotal bloc that we observe, Bloc 8, is constituted of Republicans while the State Assembly's majority is held by the Democratic party. Indeed, we find that the mutual information between the vote of Bloc 8 with the majority vote across the blocs is $I_8 = 0.96$ bits versus that of a Democratic Bloc 1, $I_1 = 0.17$ bits. This measure of correlation indicates that this Republican bloc is, like a median, highly predictive of the majority outcome across all of these blocs and, additionally, is pivotal.[6]

### 4.3.6. US House of Representatives and Senate
Data were obtained from Voteview [16]. We analyse the 80–113th Senate sessions and the 80–115th House sessions. The 80th congressional session started in 1947 and each session lasts 2 years.

For filtering voters and coarse-graining, we used the same procedure as specified for the CA Assembly above.

Data accessibility. This article has no additional data.

Authors' contributions. E.D.L. and D.M.K. conceived of the research project. E.D.L. and M.J.B. performed the computational analysis, and P.H.G. additionally helped with further calculations. E.D.L. drafted the manuscript and it was revised and reviewed by all authors. All authors gave final approval for publication and agree to be held accountable for the work performed therein.

Competing interests. We declare we have no competing interest.

Funding. E.D.L. acknowledges a dissertation grant from the Dirksen Congressional Center, an NSF GRFP under grant no. DGE-1650441, and the Omega Miller Program at the Santa Fe Institute. D.M.K. and M.J.B. thank Illinois Institute of Technology Chicago-Kent College of Law for support of this project. We thank SFI Science for covering publication costs.

Acknowledgements. We thank Katherine Quinn, Guru Khalsa, Bryan Daniels, Jess Flack and David Krakauer for helpful conversations. We thank Gavin Hall for help with the Twitter data.

## Endnotes

[1]Since the recovered eigenvector is arbitrary with respect to sign, we could just as well consider the negative eigenvector that would reverse the sign but preserve the magnitude of the elements.

[2]W. Matthews, D. Fabe and A. Bryner rotated as Chief Justice during the period 1997–2009 and W. Carpeneti from 2009 to 2012 (following the period of analysis).

[3]We are careful to point out that the ensemble of votes for political systems already include such effects so it is important to distinguish between endogenous and exogenous factors.

[4]For example, manipulation of single neurons is possible by electrical stimulation, optogenetic techniques or chemical stimulation, all ways of enacting the localized perturbations of neural 'votes' [31]. Analogously, gene expression might be perturbed by switching genes on and off or by adding protein directly to simulate changed expression levels [52].

[5]Another way to describe equation (4.1) is as the linear combination of the linear response functions of every pairwise correlation to a change in the corresponding coupling, also known as the susceptibility.

[6]Examples of blocs that are predictive of outcome but not pivotal include Bloc 3 ($I_3 = 0.96$ bits but $\tilde{\lambda}_3 = 0.02$) and A.K. and S.O. on SCOTUS [21].